\documentstyle[emulateapj,psfig,apjfonts]{article}
\def\pup{Puppis~A}
\def\rxj{RX~J0822--4300}
\def\etal{{\rm et~al.\ }}

\def\kms{km~s$^{-1}$}

\lefthead{GAENSLER, BOCK AND STAPPERS}
\righthead{NON-DETECTION OF A PULSAR-POWERED NEBULA IN PUPPIS~A}

\begin{document}
\title{Non-detection of a pulsar-powered nebula in Puppis~A, and
implications for the nature of the radio-quiet neutron star \rxj}
\submitted{(Accepted to ApJ Letters)}
\author{B. M. Gaensler\altaffilmark{1,4}, D. C.-J. Bock\altaffilmark{2}
and B. W. Stappers\altaffilmark{3}}

\altaffiltext{1}{Center for Space Research, Massachusetts Institute of
Technology, 70 Vassar Street, Cambridge, MA 02139; bmg@space.mit.edu}
\altaffiltext{2}{Radio Astronomy Laboratory, University of California at
Berkeley, 601 Campbell Hall, Berkeley, CA 94720; dbock@astro.berkeley.edu}
\altaffiltext{3}{Astronomical Institute ``Anton Pannekoek'', Kruislaan
403, 1098 SJ Amsterdam, The Netherlands; bws@astro.uva.nl}
\altaffiltext{4}{Hubble Fellow}

\begin{abstract}

We report on a deep radio search for a pulsar wind nebula associated
with the radio-quiet neutron star \rxj\ in the supernova remnant \pup.
The well-determined properties of \pup\ allow us to constrain the size of
any nebula to less than 30~arcsec; however we find no evidence for such
a source on any spatial scale up to 30~arcmin.  These non-detections
result in an upper limit on the radio luminosity of any pulsar-powered
nebula which is three orders of magnitude below what would be expected
if \rxj\ was an energetic young radio pulsar beaming away from us, and
cast doubt on a recent claim of X-ray pulsations from this source. The
lack of a radio nebula leads us to conclude that \rxj\ has properties very
different from most young radio pulsars, and that it represents a distinct
population which may be as numerous, or even more so, than radio pulsars.

\end{abstract}


\section{Introduction}
\label{sec_intro}

While the vast majority of neutron stars so far discovered are seen as
radio pulsars, there are also a small but increasing number of neutron
stars which have very different observational properties.
Approximately half these sources are soft $\gamma$-ray repeaters (SGRs)
or anomalous X-ray pulsars (AXPs), both of which show pulsed X-rays at
long periods ($P\sim10$~s) (e.g.\ \cite{mer99}). The remaining sources
are grouped together as ``radio-quiet neutron stars'' (RQNS; \cite{cbt96};
\cite{bj98}), most of which are characterized by unpulsed thermal X-ray
emission at a temperature of a few million degrees, a complete lack of
radio emission, and very high X-ray to optical ratios. Many
of these sources
have been associated with supernova remnants (SNRs), and are thus
probably quite young ($\la$20~kyr) objects.

The AXPs and SGRs are quite distinct from radio pulsars in their
properties, and are believed to be either ``magnetars'' (neutron stars
with magnetic fields $B\ga10^{14}$~G; \cite{td96b}) or exotic accreting
systems (e.g.\ \cite{vtv95}); however, an interpretation for the RQNS is
less clear.  Brazier \& Johnston (1999\nocite{bj98}) argue that RQNS are
energetic young radio pulsars like the Crab pulsar, but whose beams do not
cross our line of sight.  However, Vasisht \etal\ (1997\nocite{vka+97})
and Frail (1998\nocite{fra98}) propose that RQNS are neutron stars with
large initial periods ($P_0 \ga 0.5$~s) and/or high magnetic fields
($B \ga 10^{14}$~G) and are thus possibly related to the SGRs and AXPs,
while Geppert, Page \& Zannias (1999\nocite{gpz99}) suggest that 
they may rather be fast-spinning but weakly-magnetized sources.

One way to distinguish between all these possibilities would be
to detect pulsations from a RQNS. The period and
period derivative of the source could then be used
to infer a surface magnetic field (\cite{mt77}), while if
the RQNS can also be associated with a SNR,
an independent age determination for the latter can be used
to estimate an initial period for the neutron star (e.g. \cite{rey85}).

The RQNS \rxj\ (\cite{pbw96}) is near the center of and is almost
certainly associated with the young ($<$5000~yr; \cite{wtki88};
\cite{adp91}) and nearby (2.2~kpc; \cite{rdga95}) supernova remnant
\pup\ (G260.4--3.3).  Recently, Pavlov, Zavlin \& Tr\"{u}mper
(1999\nocite{pzt99}; hereafter PZT99) and Zavlin, Tr\"{u}mper \& Pavlov
(1999\nocite{ztp99}; hereafter ZTP99) have analyzed two archival {\em
ROSAT}\ datasets on \rxj, separated by 4.6~yr.  In each dataset they find
evidence for weak pulsations, the periods
of which are slightly different as would be expected for pulsar
spin-down.  The resulting period, $P=75.5$~ms, and period derivative,
$\dot{P}=1.49\times10^{-13}$~s~s$^{-1}$, when combined with the age
of the SNR, imply a dipole magnetic field $B = 3.4\times10^{12}$~G,
a spin-down luminosity $\dot{E} = 1.4\times10^{37}$~erg~s$^{-1}$ and an
initial period $P_0 \approx 55$~ms, all of which (despite
the radio-quiet nature of the source) are properties typical
of a young energetic radio pulsar associated with a SNR.

However energetic young pulsars in SNRs have some unmistakable
signatures. {\em Every}\ young ($\la$20~kyr) pulsar located within the
confines of a SNR powers an observable pulsar wind nebula (PWN) --- a
filled-center synchrotron source resulting from the confinement of the
relativistic pulsar wind by external pressure.  Thus a simple test to
determine if \rxj\ is indeed an energetic young pulsar, as argued by
Brazier \& Johnston (1999\nocite{bj98}) and as implied by the detection
of pulsations by PZT99 and ZTP99, is to see if it has an associated PWN.
At radio wavelengths, existing data (e.g.\ \cite{adp+90}; \cite{dbwg91})
let us put no useful constraints on the presence or absence of a PWN
associated with \rxj. This is because these observations were carried
out at relatively low frequencies (where \pup\ is brightest) and low
spatial resolution, resulting in a great deal of confusing emission at
the position of \rxj\ from both the SNR shell and from diffuse internal
emission. We have therefore carried out new observations towards \rxj,
at higher frequency and spatial resolution than previous measurements,
aimed at searching for a radio PWN associated with \rxj\ and thus
determining whether its properties are consistent with it being a
young pulsar.  Our observations are described in \S\ref{sec_obs}, while
in \S\ref{sec_results} we demonstrate the absence of any radio PWN
at the position of \rxj\ and quantify the consequent upper limits. In
\S\ref{sec_discuss} we argue that this non-detection implies that \rxj\
must have properties very different from the young energetic pulsars
which do power observable PWN.

\section{Observations}
\label{sec_obs}

Radio observations towards \rxj\ were made with the Australia
Telescope Compact Array (ATCA; \cite{fbw92}) in its 0.750D configuration on 
1999 July 24/25. In this configuration, the array contains
ten baselines in the range 31~m to 719~m, and another five
baselines in the range 3750~m to 4469~m. Since these two
sets of baselines cannot be easily combined in
a single image, this effectively
results in two sets of data: one appropriate for imaging 
extended structure on a wide range of scales (a ``large scale'' image), 
and another sensitive only to a narrow range of spatial scales, but
at much higher spatial resolution (a ``small scale'' image).

Two separate observations were made, each of duration 12~h.  In the
first, data were collected at frequencies of 1.4 and 2.5~GHz, while in
the second, a single observation was made centered at 4.8~GHz.  At 1.4
and 2.5~GHz the bandwidth was 128~MHz, while at 4.8~GHz the bandwidth
was 256~MHz, in all cases divided into 4-MHz channels.

Observations at 1.4/2.5~GHz consisted of a two-point mosaic
with mean position centered on \rxj; at 4.8~GHz, observations
consisted of a single pointing, offset 2.5~arcmin to the west
of \rxj\ so as to avoid sidelobe contamination  
from the nearby bright source PMN~J0820--4259.
Amplitudes were calibrated by observations of PKS~B1934--638,
assuming flux densities of 14.9, 11.1 and 5.7~Jy at 1.4, 2.5 and 4.8~GHz
respectively. Instrumental gains and polarization were determined
using regular observations of PKS~B0823--500. 

\section{Analysis and Results}
\label{sec_results}

After standard editing and calibration using the {\tt MIRIAD}\ package,
total intensity images were formed at each frequency, using a
multi-frequency synthesis approach to both improve
the $u-v$ coverage and minimize the effects
of bandwidth smearing (\cite{sw94}). 
At 1.4 and 2.5~GHz, mosaic images were
formed using maximum entropy deconvolution, both pointings being
deconvolved simultaneously (\cite{ssb96}). The 4.8-GHz image was deconvolved
using the {\tt CLEAN}\ algorithm. At each frequency, both  ``large scale''
and ``small scale'' images were formed. At 2.5~GHz, the
two shortest baselines sample emission from
the SNR which fills the entire field of view of the ``large scale'' image 
and prevents it from being successfully deconvolved; 
therefore at this frequency these two baselines were not used when 
forming the ``large scale'' image.

In all cases, deconvolution was constrained to only act on specific
regions of the image, namely the shell component of the SNR (defined
using the 0.8~GHz MOST image of \cite{btg98}) and background point
sources outside the SNR.  
The resulting model was subtracted from the $u-v$ data
to produce a dataset which contained visibilities corresponding only to
emission from the interior of the SNR. This dataset was then imaged,
deconvolved, smoothed with a gaussian restoring beam, and 
corrected for the mean primary beam response of the ATCA antennas.

\begin{figure*}[htb]
\centerline{\psfig{file=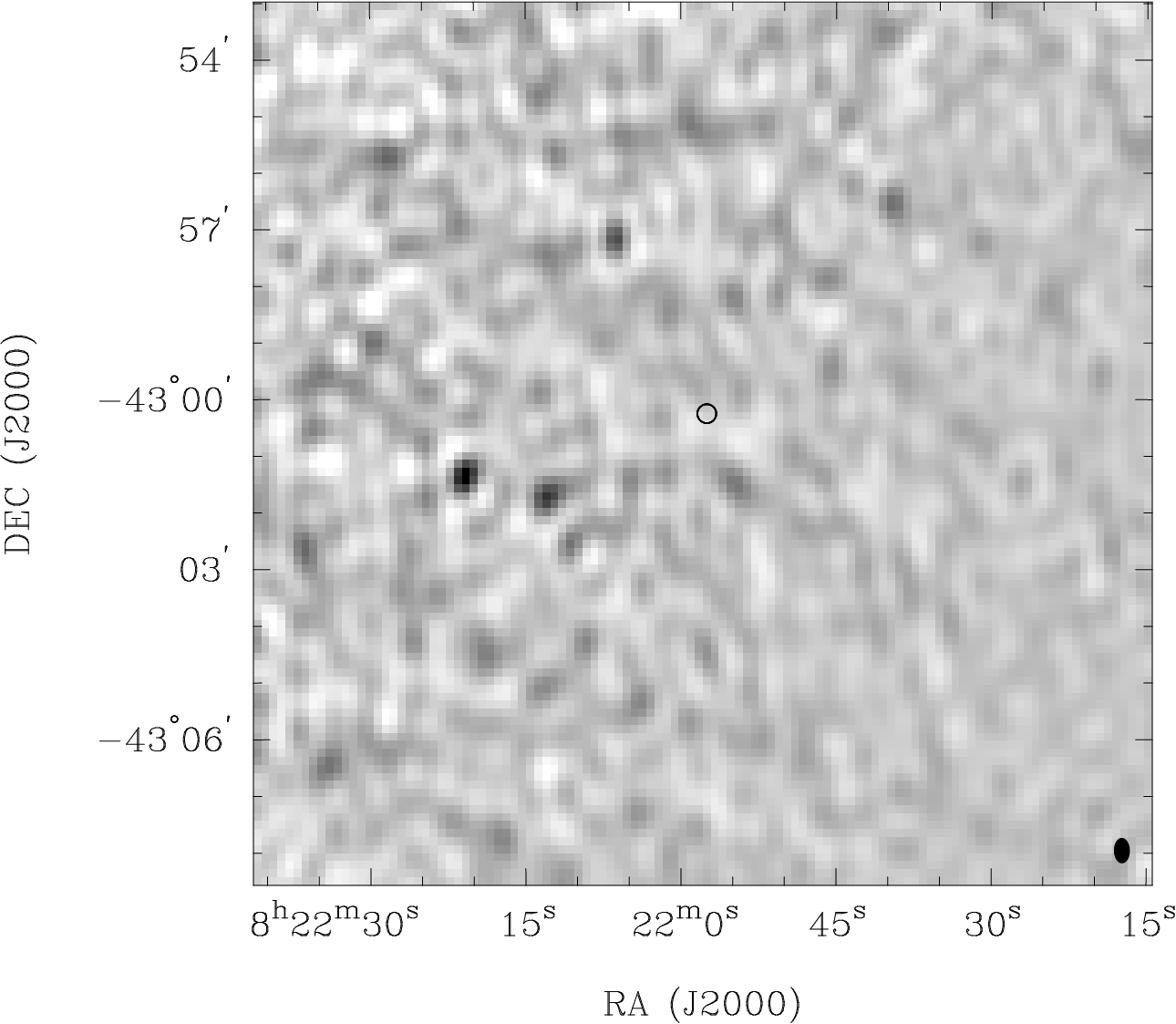,width=12cm,angle=270}}
\caption{``Large scale'' ATCA image of the region surrounding
\rxj, at a frequency of 2.5~GHz. Emission from the shell
of \pup\ and from bright point sources exterior to the SNR
has been subtracted. The circle marks the position of \rxj\
(\protect\cite{pbw96}) --- the positional uncertainty is half
the radius of the circle. The greyscale ranges from
0.0 to 2.4~mJy~beam$^{-1}$ at a resolution of
$27''\times17''$ (indicated at lower right); the rms
noise is 0.2~mJy~beam$^{-1}$.}
\label{fig_puppis}
\end{figure*}

In all six images (1.4, 2.5 and 4.8~GHz; ``small scale'' and ``large
scale'') no radio emission could be seen at or around the position of
\rxj; one such image is shown in Fig~\ref{fig_puppis}. To quantify these
non-detections we performed a series of simulations, in each of which
we modeled the appearance of a PWN by using a circular disk of a given
surface brightness and radius, centered on \rxj. This simple morphology is
a reasonable approximation to observed PWN, most of which are centered on
their associated pulsar with approximately constant surface brightness
across their extent.  In each simulation, the Fourier transform of
this disk was added to the $u-v$ data from which the shell emission
had been subtracted, and the imaging and deconvolution process was then
repeated.\footnote{The increase in antenna temperature resulting from the
flux of the disk is negligible in all cases.}  For a given diameter, we
increased the brightness of the simulated disk until it could clearly be
distinguished from the underlying noise (this criterion corresponds to a
$\sim5\sigma$ detection for small diameters, but is closer to $3\sigma$
for larger sources).  We were thus able to quantify the sensitivity of
the data to a PWN as a function of its size, incorporating effects due
to non-Gaussian noise in the image, unrelated background sources and
the limited range of spatial scales sampled by the interferometer.

The results of these simulations are shown in Fig~\ref{fig_sens}. At
each frequency, the sensitivity curve consists of four regimes. At the
smallest scales, each curve is essentially flat, corresponding to the
sensitivity of the ``small scale'' image to an unresolved source. The
curve then increases approximately as $S_{\rm min} \propto D^2$, as
expected for an extended source (cf. Fig 2 of
\cite{gsf+00}); slight deviations
from this relation are due to the effects of noise-fluctuations in the
data.  At a certain scale the sensitivity of the ``large scale'' image
to an unresolved source becomes better than that of the ''small scale''
image to a resolved source, and the curve becomes flat once more.
Finally, at scales which are resolved by the ``large scale'' image, the
curve once again increases proportional to $D^2$. The curve at each
frequency terminates at the largest scale detectable by the
interferometer; note that the 2.4~GHz curve ends prematurely because
the two shortest baselines were not used, as discussed above.

\section{Discussion and Conclusions}
\label{sec_discuss}

\begin{figure*}[htb]
\centerline{\psfig{file=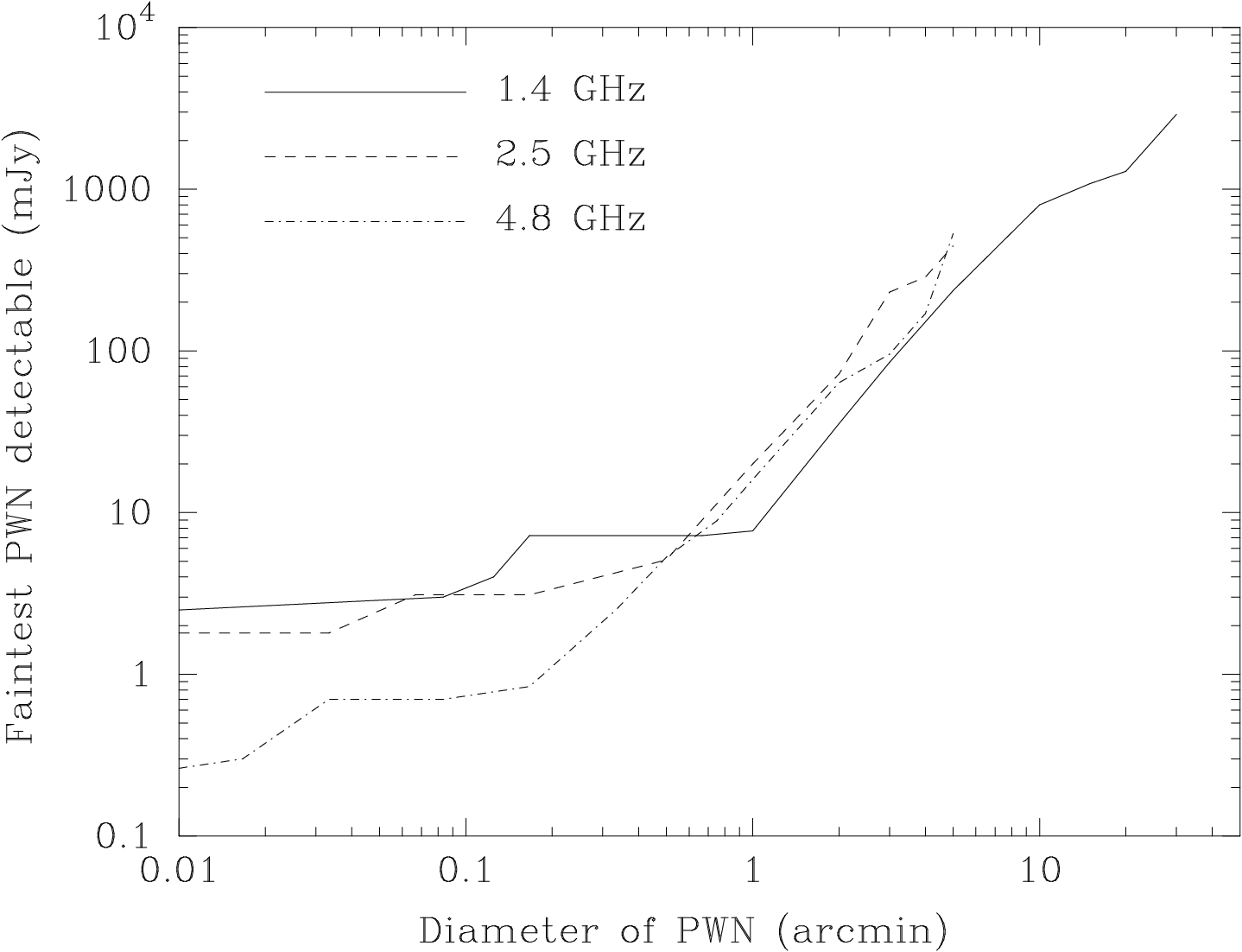,width=12cm,angle=270}}
\caption{Minimum detectable flux density of a pulsar wind nebula
at the position of \rxj\ at 1.4, 2.5 and 4.8~GHz, as
a function of size.}
\label{fig_sens}
\end{figure*}

To determine whether the limits derived in Fig~\ref{fig_sens} are
constraining, we need to determine an expected size and flux density
for a PWN associated with \rxj.  A PWN will expand until the pulsar
wind comes into pressure equilibrium with the external pressure.  As
discussed by Gaensler \etal\ (2000\nocite{gsf+00}), there are two
possible sources for this pressure: either the external gas pressure,
$p_{\rm gas}=nkT$, or the ram pressure produced by the pulsar's motion,
$p_{\rm ram}=\rho V^2$, where $n$ and $\rho$ are the number and mass
density respectively of the ambient medium, $V$
is the velocity of the pulsar, and $T$ is the temperature of ambient
gas.

By modeling the infrared emission from \pup, Arendt, Dwek \& Petre
(1991\nocite{adp91}) derive parameters for the gas interior to the SNR
(into which the PWN is expanding) of $n = 1.3$~cm$^{-3}$ and $T = 3-6
\times 10^6$~K; similar values are derived from X-ray spectroscopy
(e.g. \cite{wcc+81}; \cite{bngn94}). This corresponds to a pressure
$p_{\rm gas} = 0.05-0.11$~nPa which, if equated with the pressure
$\dot{E}/4\pi r^2 c$ from the pulsar wind (where $r$ is the radius of
the PWN and $\dot{E} = 1.4\times 10^{37}$~erg~s$^{-1}$), results in a
PWN of diameter 11--16~arcsec at a distance 2.2~kpc.

For the same number density as used above, and assuming the
ambient gas to be composed only of atomic hydrogen, we
find that $p_{\rm ram} =
2.2V_3^2$~nPa, where $V = 10^3 V_3$~\kms. Balancing
this pressure with that from the pulsar wind 
(e.g. \cite{gsf+00}) results in a PWN of angular diameter 
$\sim4V_3^{-1}$~arcsec.

Since the mean velocity of the pulsar population is $V_3 \approx 0.38$
(\cite{cc98}), and in fact in this particular instance the offset of \rxj\
from the dynamical center of the SNR argues that $V_3 > 1$ (\cite{pbw96}),
it is likely that $p_{\rm ram} > p_{\rm gas}$, and that the smaller of the
two sizes we have just estimated, corresponding to a bow-shock morphology,
should apply. We note that in such a case, it is still reasonable to
model the PWN as a circular disk, since in observed bow-shock nebulae
most of the emission is concentrated close to the head of the nebula.
In any case, regardless of the dominant source of confining pressure,
the expected extent of a PWN powered by \rxj\ is small. Although we
believe the sizes derived above to be robust, we conservatively adopt
a maximum angular size for any PWN of 30~arcsec to take into account
possible uncertainties in $V$, $n$, $T$, or the distance to the source.
From Fig~\ref{fig_sens}, it can be seen that at all three frequencies,
the upper limit on the flux density for such a source is $\sim$7~mJy.

Assuming a typical PWN spectral index of $\alpha = -0.3$ ($S_\nu \propto
\nu^{\alpha}$), an upper limit of 7~mJy at 1.4~GHz corresponds to a
broad-band radio luminosity (integrated between 10~MHz and 100~GHz)
of $L_R = 2 \times 10^{30}$~erg~s$^{-1}$.  Defining $\epsilon \equiv
L_R/\dot{E}$ to be the ratio between a PWN's broad-band radio luminosity
and its spin-down luminosity, we find that for $\dot{E} = 1.4\times
10^{37}$~erg~s$^{-1}$ as reported by PZT99 and ZTP99, we can derive
an upper limit $\epsilon < 10^{-7}$. This is a more stringent limit
on $\epsilon$ than has been derived for almost any other pulsar
(cf.\ \cite{fs97}; \cite{gsf+00}). In particular, this upper limit
is sharply at odds with the values of $\epsilon$ seen for other young
($\la 20$~kyr) pulsars, almost all of which produce radio PWN or have
upper limits consistent with $\epsilon \ge 10^{-4}$ (\cite{fs97};
\cite{gsf+00}). The glaring exception to this is PSR~B1509--58 in
the SNR~G320.4--1.2 (MSH~15--5{\em 2}), 
which powers an X-ray PWN but for which no
radio PWN has yet been detected (\cite{gbm+98}).  However, this can be
understood in terms of the low ambient density ($n < 0.01$~cm$^{-3}$),
which results in severe adiabatic losses and a consequently underluminous
radio PWN (\cite{bha90}). This condition is not satisfied for \rxj, and
so cannot be considered as a possible explanation for the non-detection
of a PWN.\footnote{Furthermore, PSR~B1509--58 powers a bright X-ray PWN
(e.g.\ \cite{shss84}; \cite{bb97}), while no X-ray nebula is seen
around \rxj\ (\cite{psz00}).} We thus find that any PWN in \pup\
has a radio luminosity three orders of magnitude fainter than expected
for the spin parameters derived by PZT99 and ZTP99.

Nevertheless, if we assume, as Brazier \& Johnston (1999\nocite{bj98})
have argued, that \rxj\ is a rotation-powered pulsar, what spin
parameters can we infer for it? If we require that $\epsilon \ge
10^{-4}$ as seen for other young pulsars, the maximum value of $\dot{E}
\equiv 4\pi^2 I \dot{P}/P^3$ which is consistent with our non-detection
of a radio PWN is $\sim10^{33}$~erg~s$^{-1}$. Meanwhile, it is unlikely
that the characteristic age, $\tau \equiv P/2\dot{P}$, of the pulsar is
more than 50~kyr, $\sim$10 times the true age of the system.  These
upper limits on $\dot{E}$ and $\tau$ correspond to lower limits
$P>3.5$~s, $\dot{P} > 1.1 \times 10^{-12}$~s~s$^{-1}$ and $B>6.4 \times
10^{13}$~G, parameters which are very similar to those seen
for the SGRs/AXPs (\cite{kcs99}) and for the young radio
pulsar PSR~J1814--1744 (\cite{pkc00}; \cite{ckl+00}), but quite
different than those of other young pulsars in SNRs, for which
typically $P<0.2$~s, $\dot{E} > 10^{36}$~erg~s$^{-1}$ and $B\approx
10^{12}$~G.

Whether \rxj\ indeed has a long initial period and high magnetic
field, or has some other properties such that it does not produce a
detectable radio nebula or radio pulsations, the lack of a PWN
around this source (and around other RQNS such as 1E~1207.4--5209 in
the SNR~G296.5+10.0; \cite{mbc96}; \cite{gdg+00}), argues that at least
some RQNS have drastically different properties from young
radio pulsars.

Brazier \& Johnston (1999\nocite{bj98}) list six RQNS which are younger
than 20~kyr and nearer than 3.5~kpc. Excluding two RQNS from their list
which do power PWN and thus may well be radio pulsars beaming away from
us, but including the recently-discovered RQNS in the young and nearby
SNR~Cassiopeia~A (\cite{tan99}; \cite{pza+00}; \cite{cph+00}), this
implies a Galactic birth-rate for such sources of at least once every
$\sim$200 years, comparable to or even in excess of the birth-rate for
radio pulsars (e.g.\ \cite{lml+98}). Radio-quiet neutron stars thus
point to the possibility that pulsars like the Crab are not the most
common manifestation of neutron star.

\begin{acknowledgements}

We thank Froney Crawford for assistance with the observations,
and George Pavlov and Ulrich Geppert for helpful suggestions.
The Australia Telescope is funded by the Commonwealth of Australia for
operation as a National Facility managed by CSIRO.  B.M.G.  acknowledges
the support of NASA through Hubble Fellowship grant HF-01107.01-98A
awarded by the Space Telescope Science Institute, which is operated
by the Association of Universities for Research in Astronomy, Inc.,
for NASA under contract NAS 5--26555. B.W.S. is supported by NWO Spinoza
grant 08-0 to E.P.J. van den Heuvel.

\end{acknowledgements}


\bibliographystyle{apj1}
\bibliography{journals,modrefs,psrrefs,crossrefs}

\clearpage

\vspace{1cm}

\end{document}